\documentclass{article}
\usepackage[affil-it]{authblk}
\usepackage{ragged2e}
\usepackage{indentfirst}

\usepackage{xcolor}
\usepackage{hyperref}

\usepackage{PRIMEarxiv}

\usepackage[utf8]{inputenc} 
\usepackage[T1]{fontenc}   
\usepackage{hyperref}      
\usepackage{url}           
\usepackage{booktabs}      
\usepackage{amsfonts}      
\usepackage{nicefrac}     
\usepackage{microtype}     
\usepackage{lipsum}
\usepackage{fancyhdr}      
\usepackage{graphicx}      
\graphicspath{{media/}}     
\title{Critical heat flux diagnosis using conditional generative adversarial networks}
\author{UngJin Na\textsuperscript{a}, Moonhee Choi\textsuperscript{c}, HangJin Jo\textsuperscript{a, b,}\thanks{Corresponding author, jhj04@postech.ac.kr}}
\begin{document}

\maketitle
\begin{center}
\textit{\small\textsuperscript{a}Department of Mechanical Engineering, Pohang University of Science and Technology (POSTECH),} \\
\textit{\small 77 Cheongam-Ro, Nam-gu, Pohang, Gyeongbuk 37673, Korea} \\
\textit{\small\textsuperscript{b}Division of Advanced Nuclear Engineering, Pohang University of Science and Technology (POSTECH),} \\
\textit{\small 77 Cheongam-Ro, Nam-gu, Pohang, Gyeongbuk 37673, Korea} \\
\textit{\small\textsuperscript{c}Department of Nuclear Engineering, Seoul National University,} \\
\textit{\small 1 Gwanak-ro, Gwanak-gu, Seoul, 08826, Korea}
\end{center}

\vskip 0.23in%

\begin{abstract}
The critical heat flux (CHF) is an essential safety boundary in boiling heat transfer processes employed in high heat flux thermal-hydraulic systems. Identifying CHF is vital for preventing equipment damage and ensuring overall system safety, yet it is challenging due to the complexity of the phenomena. For an in-depth understanding of the complicated phenomena, various methodologies have been devised, but the acquisition of high-resolution data is limited by the substantial resource consumption required. This study presents a data-driven, image-to-image translation method for reconstructing thermal data of a boiling system at CHF using conditional generative adversarial networks (cGANs). The supervised learning process relies on paired images, which include total reflection visualizations and infrared thermometry measurements obtained from flow boiling experiments. Our proposed approach has the potential to not only provide evidence connecting phase interface dynamics with thermal distribution but also to simplify the laborious and time-consuming experimental setup and data-reduction procedures associated with infrared thermal imaging, thereby providing an effective solution for CHF diagnosis.

\end{abstract}


\section{Introduction}

The critical heat flux (CHF) represents the maximum heat flux in the nucleate boiling process, marking an abrupt increase in surface temperature. As a crucial factor in high heat-flux systems to ensure safe operation and prevent system damage, CHF diagnosis has been extensively researched, leading to the development of various mechanistic models explaining the triggering mechanisms of CHF \cite{weisman1983prediction}\cite{haramura1983new}\cite{lee1988mechanistic}\cite{galloway1993chf}. Among these models — such as the hydrodynamic instability model, macrolayer dryout model, and interfacial lift-off model — the hot/dry spot model suggests that irreversible dry patch formation leads to increasing temperature, resulting in the postulation that the development of the irreversible dry spot's temperature hinders the wetting of the heated surface by the supplied liquid \cite{liang2018pool}. The dry patch is first generated at high heat flux, then coalesces and expands again under the remnant bubble to trigger CHF \cite{jeon2022observation}.

To validate and improve such models, visual observation methods have been developed \cite{chu2014observation}\cite{kim2018mechanism}. Total reflection visualization and (TR) infrared thermometry (IR) are arguably the most important techniques for visualizing the formation of dry patches while measuring the coincidental temperature evolution of the liquid-vapor system \cite{nishio1998observation}\cite{bucci2016mechanistic}\cite{choi2020direct}. Through the methods, the behavior of the bubble structure and dry patch under flow boiling has been observed, and the hydrodynamic mechanism of the irreversible dry patch have been analyzed. Also, there have been attempts to determine CHF based on the temperature of the dry patch periphery \cite{jeon2022observation}\cite{choi2021observation}.

Besides, following recent advancements in Convolutional Neural Networks (CNNs), which excel in capturing visual information characteristics, neural networks are expected to have the potential to simplify infrared thermal imaging, as the process typically involves tedious experimental setups and extensive data reduction \cite{krizhevsky2017imagenet}. In particular, the adoption of Generative Adversarial Networks (GANs) has facilitated image generation while learning the style representation of inputs, allowing many challenges in image processing to be addressed within a common framework \cite{goodfellow2020generative}. By combining both CNNs and GANs, Deep Convolutional Generative Adversarial Networks (DCGANs) can more effectively capture characteristic visual information while stabilizing the learning process \cite{radford2015unsupervised}. Furthermore, Conditional Generative Adversarial Networks (cGANs) have enabled automatic image-to-image translation through a convolutional generator-discriminator architecture \cite{isola2017image}.

In this study, a novel approach for CHF diagnosis is proposed, utilizing a data-driven, image-to-image translation method that leverages cGANs. This approach has the potential to streamline the imaging process of the temperature field and significantly enhance the speed of extracting valuable information from large datasets and generating visual images. For the development of reliable methodologies, well-defined experimental data featuring total reflection and IR images are used, and the modified cGAN architecture is employed to generate artificial results from the original images. The comparison between experimentally measured IR and generated IR images demonstrates the capability to translate images involving physical phenomena from gray images (TR images) to RGB images (IR images). Additionally, the latent space is investigated using principal component analysis, which supports the reliability of the proposed methodology for thermal distribution analysis.

\section{Methodology}

\subsection{The data acquisition}

The datasets for this study were generated based on the flow boiling experiment, which aimed to obtain total reflection images and infrared images. The setup is as follows: The main test section consisted of a square channel, with its four sides made up of three transparent polycarbonate plates and one sapphire plate. The primary heating surface utilized in the experiment was a sapphire specimen coated with Indium Tin Oxide (ITO) due to its high transparency and low IR transmittance, allowing for the direct measurement of the heating surface temperature. A triangular prism-shaped silicone oil bath was placed at the bottom of the sapphire plate to set the incident angle for total reflection at the boiling surface. The experiment covered a nucleate boiling regime, ranging from discrete bubbles to critical heat flux conditions. The criteria for the occurrence of critical heat flux were ascertained by simultaneously observing total reflection and IR images. The HSV cameras used in the experiment were Memrecam GX-3 (color), manufactured by NAC Inc., and a FLIR X6903sc camera for IR thermal imaging to gather boiling temperature information. The details of the experiment can be found in \cite{choi2021observation}.
 
\begin{figure}[hbt!]
\centering
\includegraphics[width=0.9\textwidth]{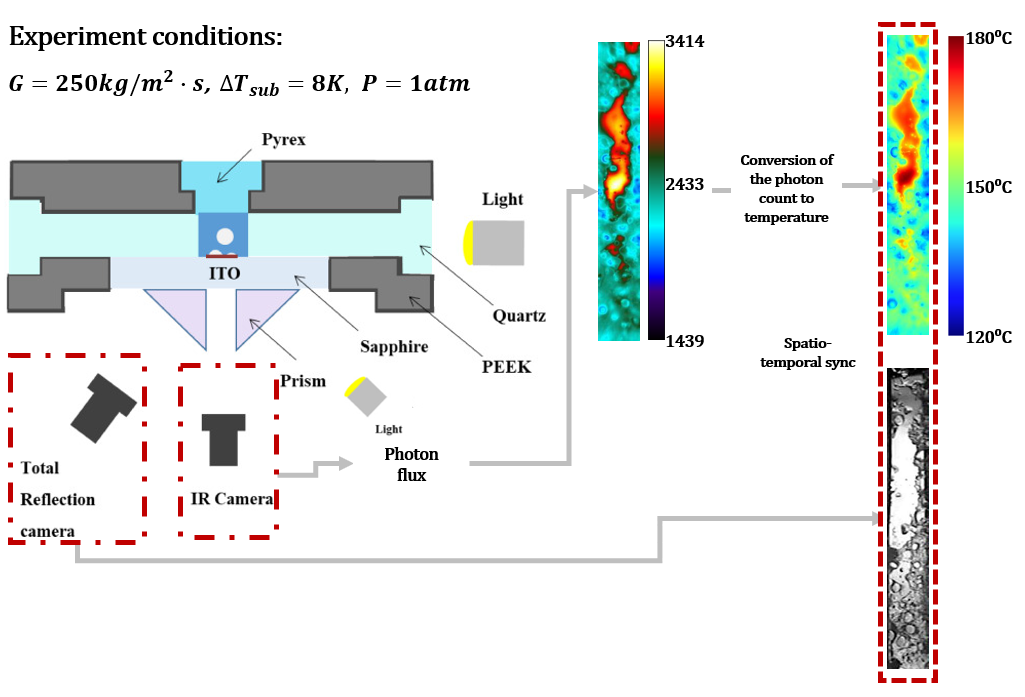} 
\caption{Schematic of visualization experiment}
\end{figure}

\subsection{cGAN implementation}

The process of translating total reflection images to infrared images can be formulated as follows:

\begin{equation}
    G : X \rightarrow Y,\quad \quad X = \{x_{i}\}_{i=1}^{N},\quad Y = \{y_{i}\}_{i=1}^{N}
\end{equation}

where $x$ and $y$ represent the total reflection image and the infrared image, respectively, and $i$ signifies the sequential index of the images. The process is addressed using a cGAN, in which the generator (G) learns to transform an input image into an output image that resembles the target. This ensures that the conditional probability of the prediction given the input approaches the conditional probability of the ground truth given the input. Meanwhile, the discriminator (D) aims to differentiate real images from generated ones.

The objective function of the cGAN, $\mathcal{L}_{\mathrm{cGAN}}(D, G)$, is formulated as follows \cite{isola2017image}:

\begin{equation}
\mathcal{L}_{\mathrm{cGAN}}(D, G) = \mathbb{E}_{y \sim p_{\mathrm{data}}(y)}[\log D(y|x)] + \mathbb{E}_{\hat{y} \sim p_{\mathrm{data}}(\hat{y})}[\log(1 - D(G(\hat{y}|x)))]
\end{equation}

The first term in the equation denotes the expectation of the log probability of the discriminator correctly classifying real images ($y$) given the condition ($x$). The second term represents the expectation of the log probability of the discriminator incorrectly classifying generated images, which are produced by the generator ($\hat{y}$) and given the condition ($x$). The cGAN is trained using the binary cross-entropy loss function, which measures the similarity between the generated output and the ground truth.

To optimize the cGAN, the following saddle-point optimization problem is solved:

\begin{equation}
G^* = \arg \min_{G} \max_{D} \left[\mathcal{L}_{\mathrm{cGAN}}(D, G) + \mathcal{L}_{L1}(G)\right]
\end{equation}

As the equations indicate, the cGAN training occurs while the generator aims to minimize the objective function, which means that the generator striving to create convincing infrared images that deceive the discriminator, and the discriminator attempting to differentiate between genuine and generated infrared images. The inclusion of the $\mathcal{L}_{L1}(G)$ term in the optimization problem helps to match the details of the ground truth thereby improving the quality of the generated images. By iteratively updating the generator and discriminator through this optimization process, this model effectively learns to generate realistic images that closely resemble the ground truth.

\begin{figure}[hbt!]
\centering
\includegraphics[width=1\textwidth]{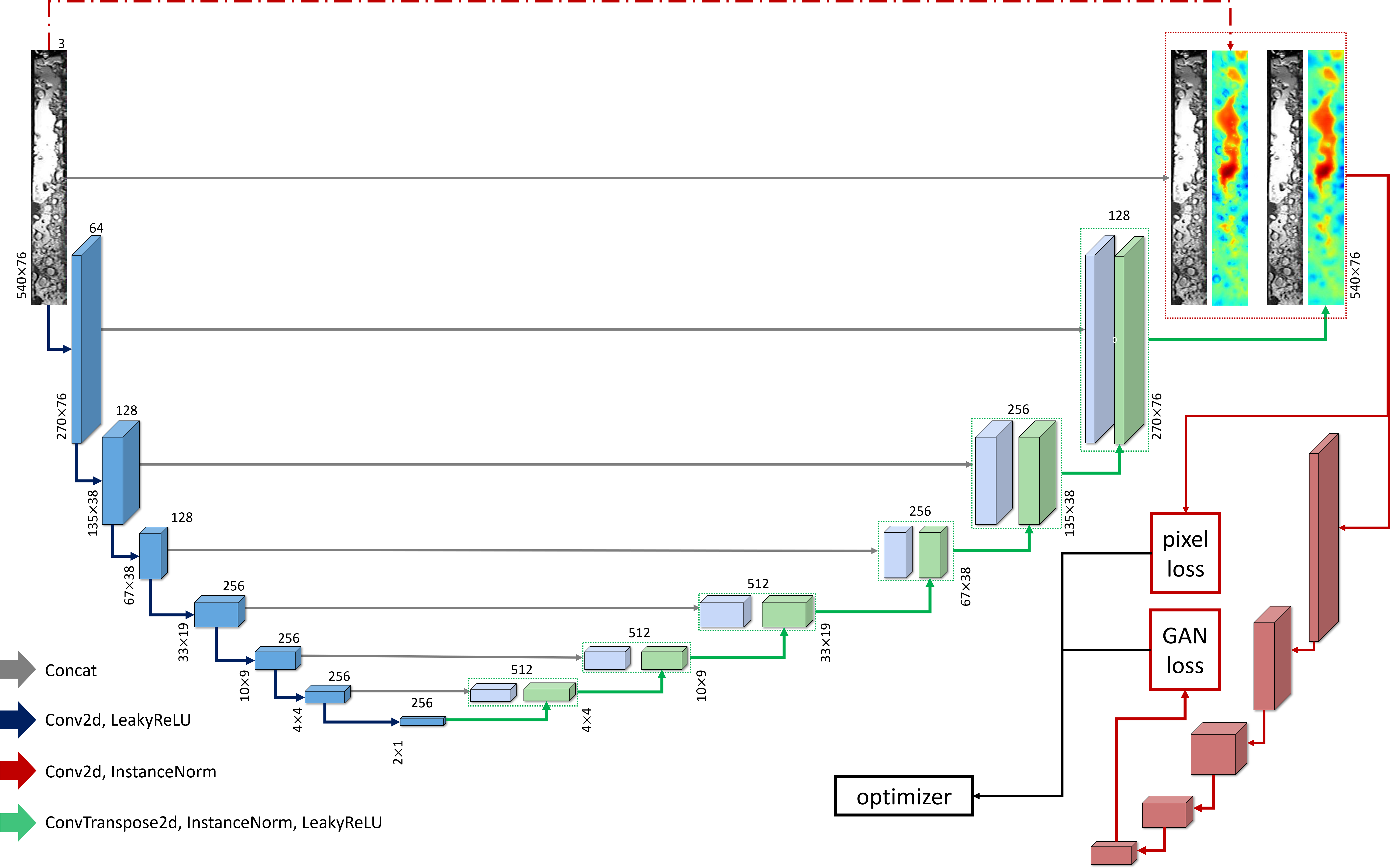}  
\caption{Illustration of the modified cGAN architecture}
\end{figure}

In this study, a modified U-Net architecture was utilized as a generative model for synthesizing images. The generator part of the architecture consists of an encoding and a decoding path. The encoding path is composed of a series of downsampling modules that capture the context and spatial information of the input image by translating the total reflection image domain \(X\) into the latent space \(z\). The decoding path is made up of a series of upsampling modules which recovers the detailed information and generates the output image by converting latent variables \(z\) to the infrared image domain \(Y\). The skip connections between corresponding layers in both paths help preserve low-level information shared between the input and output.

The discriminator part of the architecture is designed to penalize the high-frequency information structure in local image patches. The modified PatchGAN classifies each patch in an image as either real or fake and averages all responses to provide the output of D. By restricting the attention of the discriminator to each image patch, it effectively models the image as a Markov random field to capture local information.

The model is trained using paired images of optical visualization and infrared thermometry measurement results, serving as the ground truth, from the flow boiling experimental cases. Two experimental datasets were used, each consisting of 1650 images, and were randomly shuffled to separate images for the train and test datasets. The model implementation was done with PyTorch 1.12 and CUDA 11.3, on an RTX 3070 Ti graphics card in an Ubuntu 20.04 environment.

\section{Results}

\subsection{Snapshots of the generated images and dry patch behaviour}

Figure 3 presents three different moments captured in a total reflection view and temperature profile translated by the generator, along with the ground truth IR images during the CHF occurrence, denoted by TR, GAN, and IR images. The growth of dry patches and the increase in average temperature can be observed. The residual dry patch continued to grow and triggered CHF. When rewetting failure occurred, the maximum temperature continued to increase and the size of the dry patch no longer decreased.

\begin{figure}[hbt!]
\advance\leftskip3mm
\includegraphics[width=0.9\textwidth]{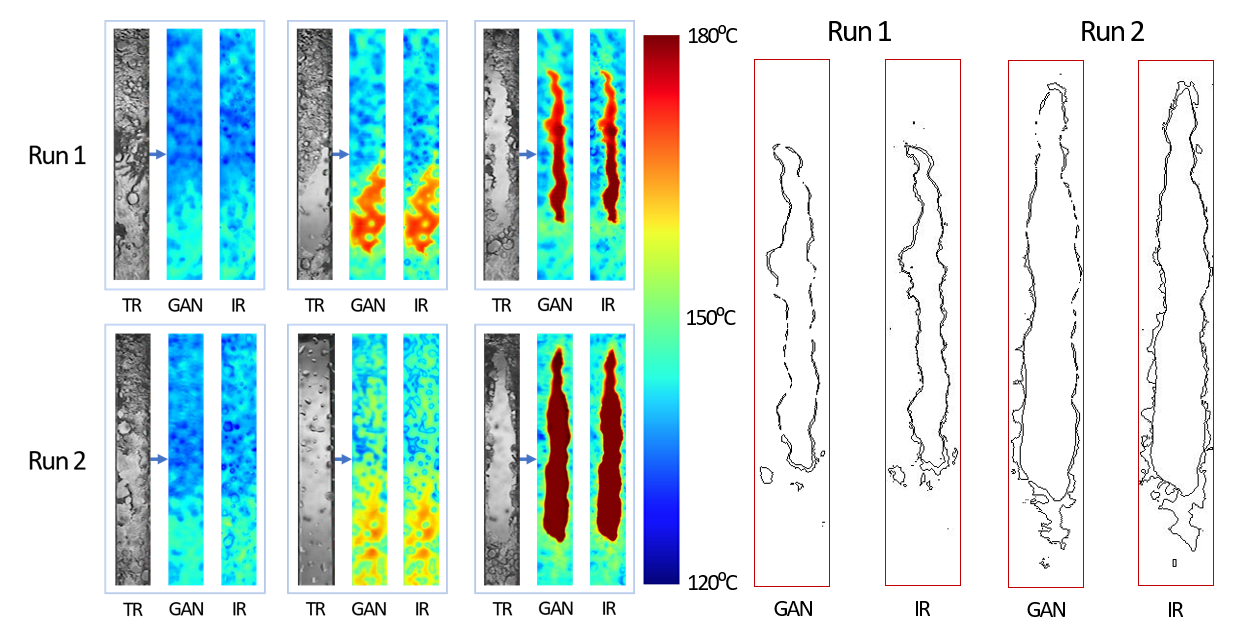}  
\caption{Snapshots of the generated images and dry patch behaviour}
\end{figure}

From the generated temperature field, a threshold of 167°C was applied to both Run 1 and Run 2, enabling the identification of the irreversible dry patches in the temperature field by presenting the contours for each of the experimental runs. Despite the differences observed between Run 1 and Run 2, the primary objective of the study was to learn the dry patch behavior, which was successfully achieved.

To quantify the accuracy of the model, PSNR and SSIM are used. The PSNR is given by:

\begin{equation}
{PSNR} = 10 \cdot \log_{10}\left(\frac{{MAX}^2_I}{{MSE}}\right) ,\ \quad  {MSE} = \frac{1}{MN} \sum_{i=1}^M \sum_{j=1}^N [I(i, j) - \hat{I}(i, j)]^2
\end{equation}

where ${MAX}_I$ is the maximum pixel value of the image, and ${MSE}$ is the Mean Squared Error between the original image ($I$) and the reconstructed image ($\hat{I}$), where $M$ and $N$ are the dimensions of the images.

The SSIM metric measures the structural similarity between two images, taking into account luminance, contrast, and structural information. The SSIM is given by \cite{wang2004image}:

\begin{equation}
{SSIM}(I, \hat{I}) = \frac{(2\mu_I \mu_{\hat{I}} + C_1)(2\sigma_{I\hat{I}} + C_2)}{(\mu_I^2 + \mu_{\hat{I}}^2 + C_1)(\sigma_I^2 + \sigma_{\hat{I}}^2 + C_2)}
\end{equation}

where $\mu_I$ and $\mu_{\hat{I}}$ are the means of the original and reconstructed images, respectively; $\sigma_I^2$ and $\sigma_{\hat{I}}^2$ are their variances, respectively; $\sigma_{I\hat{I}}$ is the covariance between the original and reconstructed images; and $C_1 = (k_1 L)^2$ and $C_2 = (k_2 L)^2$, with $L$ being 255, and $k_1 = 0.01$ and $k_2 = 0.03$.

The SSIM and PSNR values for each run are as follows:

\begin{table}[hbt!]
  \centering
  \begin{tabular}{lcc}
    \toprule
                        & \textbf{Run 1} & \textbf{Run 2} \\ 
    \midrule
    \textbf{PSNR} & 23.73     & 22.49     \\
    \textbf{SSIM} & 0.7476    & 0.7277    \\
    \bottomrule
  \end{tabular}
  \vspace*{5mm}
    \caption{Comparison of Average PSNR and SSIM for Run 1 and Run 2}
\end{table}

\subsection{Temperature field evolution}

The temperature field evolution of the output layer data for Run1 and Run2 was analyzed. The images were converted into temperature values by recreating a look-up table for the colorbar used, evaluating the model performance quantitatively. Figure 4 displays the trends of the average temperature and the maximum temperature.

\begin{figure}[hbt!]
\centering
\includegraphics[width=1\textwidth]{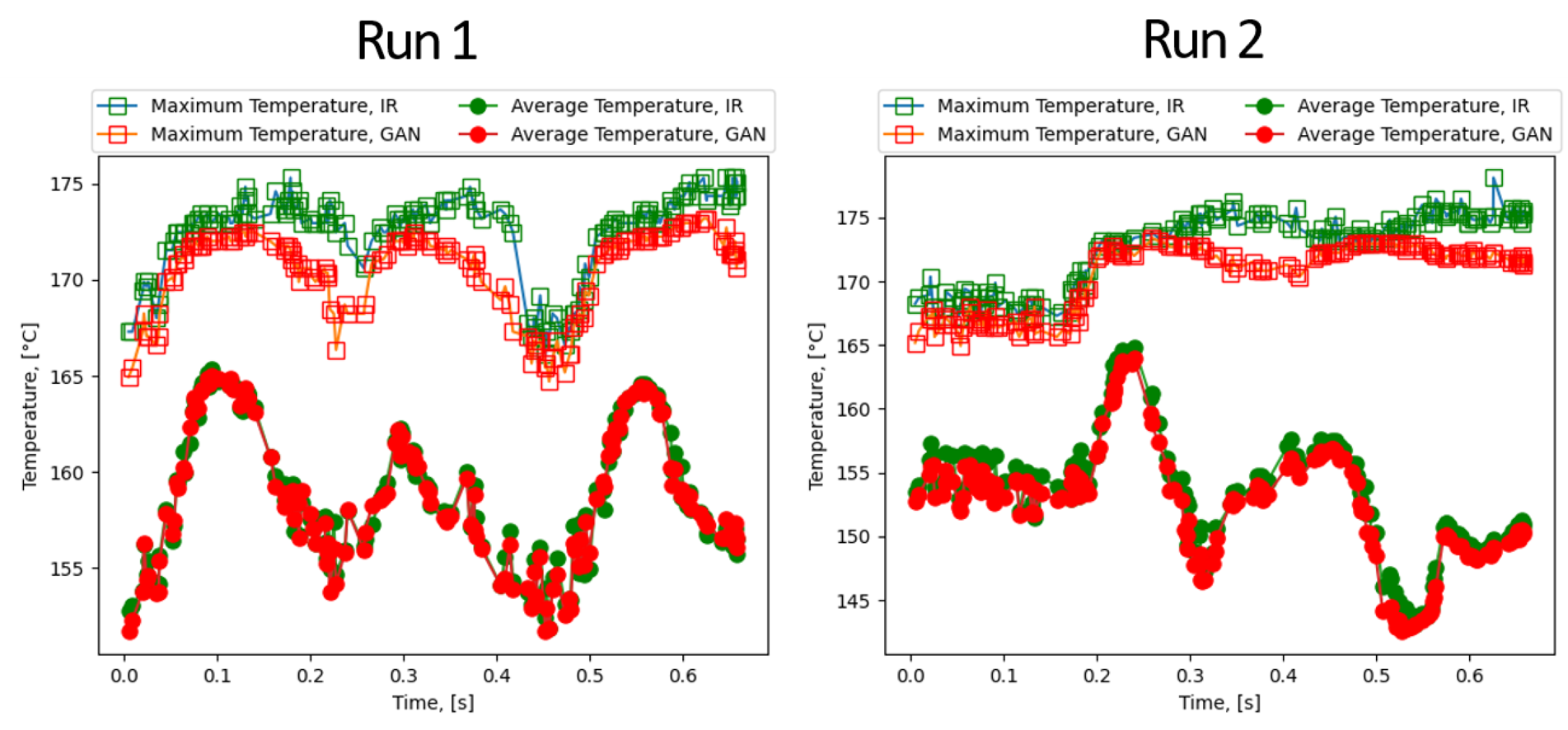} 
\caption{Temperature field evolution in the generated data}
\end{figure}

The first notable observation from the temperature field evolution is that the model demonstrates good agreement between the predicted average temperature values and actual values for both runs. This indicates that the model is capable of capturing the overall trends and patterns in the temperature field. The success in approximating the average temperature can be attributed to the use of the U-Net architecture, which allows the model to preserve spatial information through skip connections, thereby enhancing its predictive capabilities. Additionally, the maximum temperature was slightly underestimated at rates of 3.5\% and 3.3\% for Run1 and Run2, respectively. This underestimation may be attributed to the limitations of the model in capturing extreme values or localized phenomena, which could be influenced by factors such as the choice of architecture, hyperparameters, or training data. Despite this underestimation, the model's ability to provide reasonable approximations of the temperature field evolution is noteworthy, given the complexity of the problem.

\subsection{Principal Component Analysis of the latent variables}

While the previous sections presented results from the output layer, an alternative approach is required to investigate the deepest part of the generator, after the encoder portion. The encoder effectively compresses information into a dimension of $z_{i} \in \mathbb{R}^{256 \times 1 \times 2}$. To further analyze this latent representation, a principal component analysis was conducted on the encoded data and presented alongside temperature evolution, both normalized using min-max scaling within the range of [-1, 1].

\begin{figure}[hbt!]
\centering
\includegraphics[width=0.9\textwidth]{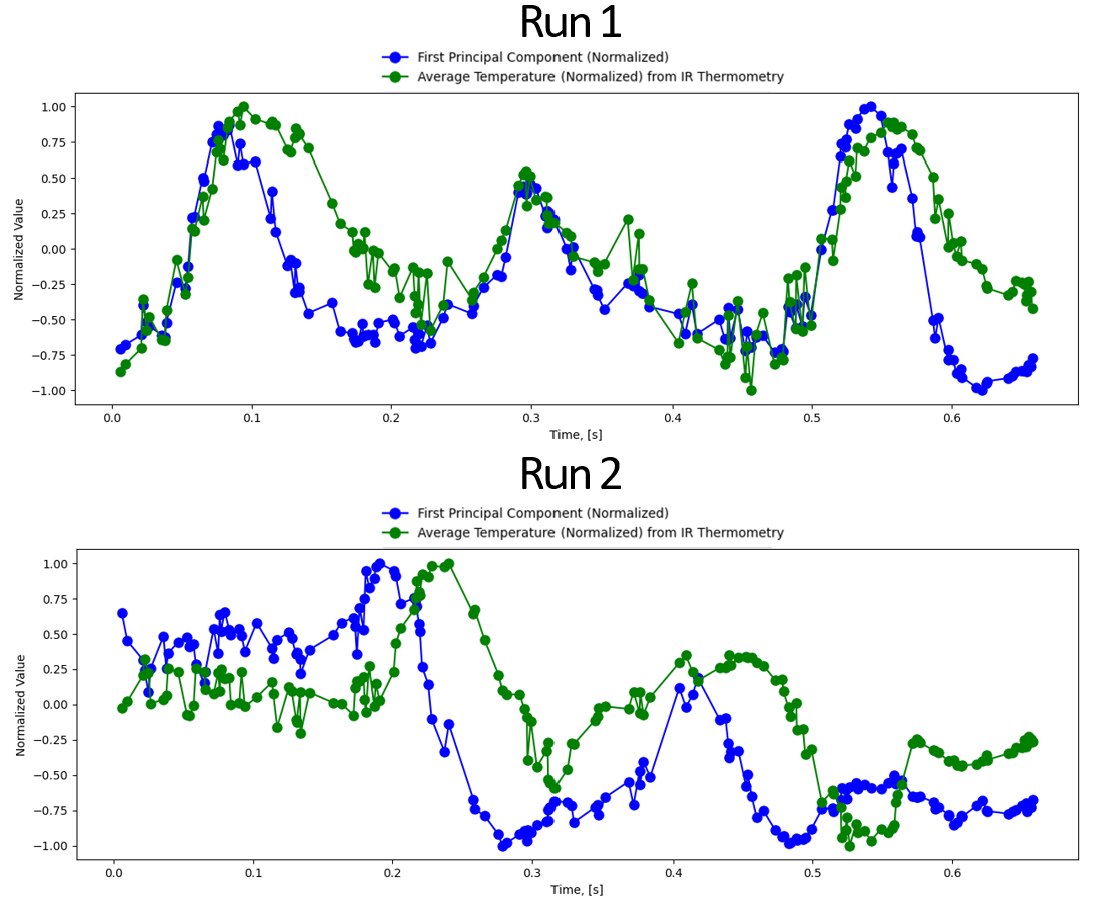}  
\caption{The first principal component and the average temperature}
\end{figure}

Through PCA, the first principal component, which is the most characteristic, was obtained. This component displayed a similar trend to the average temperature of the experimental data. In comparing Run 1 and Run 2, Run 1 appeared to better illustrate the similarity between the trends in the first principal component and the average temperature. This observation provides strong evidence that the average temperature, being the most critical element of the learning layers in the neural network, can be identified not only from the output layer but also from the latent space of the generator. This information can be utilized to examine the learning representation of the model for the experimental data.

\section{Discussion and Conclusion}

In this study, we introduced a novel approach for CHF diagnosis using a data-driven, image-to-image translation method employing conditional generative adversarial networks. This method holds the potential to streamline the temperature field generation process and significantly enhance the efficiency of extracting valuable information from large experimental datasets and making predictions in infrared thermal imaging. The temperature field evolution results demonstrated that the U-Net-based generative model provided a satisfactory estimation of the average temperature but faced challenges in predicting maximum temperature values. Furthermore, by performing a principal component analysis on the encoded data, we explored the deepest part of the generator, revealing that the average temperature, the most critical element in the learning layers of the neural network, can be identified and analyzed not only from the output layer but also from each latent space of the generator. These findings underscore the model's strengths and challenges.
It is noteworthy that the model is trained on paired input and output image examples, with the output image quality depending on the input image quality and the cGAN's ability to learn the mapping between them. If input images are of low quality or unrepresentative of the data distribution, the cGAN may generate subpar outputs that fail to accurately represent the underlying relationship between input and output. Future work could concentrate on improving the model's ability to diagnose extreme temperature values more precisely, for example, by incorporating additional features or optimizing the model's parameters.

\section*{Acknowledgments}
This work was supported by Korea Hydro \& Nuclear Power Co and Local Government (Pohang). (2023)

\bibliographystyle{unsrt}  
\bibliography{reference}

\newpage
\section*{Supplementary material: implementation of the modified cGAN architecture}

\subsection*{Generator}

The forward pass of the generator architecture involves passing the input through each of the downsampling layers, followed by the corresponding upsampling layers, with skip connections between the downsampling and upsampling layers to preserve spatial information \cite{ronneberger2015unet}. The output of the generator is produced by the final sequential layer, which applies a Tanh activation function to ensure that the pixel values of the generated image lie within the range of -1 to 1. The specific design choices, such as the number of channels, kernel size, stride, padding, normalization, and dropout rates, are as follows. 

The encoding path consists of seven downsampling layers, utilizing 2D instance normalization. Each layer is followed by a LeakyReLU with a threshold value of 0.2.

\begin{table}[hbt!]
\centering
\begin{tabular}{|c|c|c|c|c|c|c|c|}
\hline
Layer & Input Channels & Output Channels & Kernel Size & Stride & Padding & Normalization & Dropout Rate \\
\hline
1 & 3 & 64 & (4, 3) & (2, 1) & 0 & No & - \\
\hline
2 & 64 & 128 & (4, 4) & (2, 2) & 1 & Yes & - \\
\hline
3 & 128 & 128 & (5, 3) & (2, 1) & 1 & Yes & - \\
\hline
4 & 128 & 256 & (5, 4) & (2, 2) & 1 & Yes & 0.5 \\
\hline
5 & 256 & 256 & (6, 3) & (3, 2) & 0 & Yes & 0.5 \\
\hline
6 & 256 & 256 & (4, 3) & (2, 2) & 0 & Yes & 0.5 \\
\hline
7 & 256 & 256 & (3, 4) & (1, 1) & 0 & No & 0.5 \\
\hline
\end{tabular}
\end{table}

The decoding path comprises six upsampling layers and a final sequential layer, which are made up of ConvTranspose2D. Each layer is followed by a 2D instance normalization layer and a LeakyReLU with a threshold value of 0.2.

\begin{table}[hbt!]
\centering
\begin{tabular}{|c|c|c|c|c|c|c|}
\hline
Layer & Input Channels & Output Channels & Kernel Size & Stride & Padding & Dropout Rate \\
\hline
1 & 256 & 256 & (3, 4) & (1, 1) & 0 & 0.5 \\
\hline
2 & 512 & 256 & (4, 3) & (2, 2) & 0 & 0.5 \\
\hline
3 & 512 & 256 & (6, 3) & (3, 2) & 0 & 0.5 \\
\hline
4 & 512 & 128 & (5, 4) & (2, 2) & 1 & 0.5 \\
\hline
5 & 256 & 128 & (5, 3) & (2, 1) & 0 & - \\
\hline
6 & 256 & 64 & (4, 4) & (2, 2) & 1 & - \\
\hline
7 & 128 & 3 & (4, 3) & (2, 1) & 1 & - \\
\hline
\end{tabular}
\end{table}

\subsection*{Discriminator}

The discriminator architecture comprises four discriminator block layers and a convolutional layer to extract features from generated images and reduce their spatial dimensions. Each block layer is followed by a LeakyReLU with a threshold value of 0.2. The discriminator's structure is shown in the table below:

\begin{table}[hbt!]
\centering
\begin{tabular}{|c|c|c|c|c|c|}
\hline
Layer & Input Channels & Output Channels & Kernel Size & Stride & Padding \\
\hline
1 & $3 \times 2$ & 64 & (4, 3) & (2, 1) & 1 \\
\hline
2 & 64 & 128 & (4, 4) & (2, 2) & 1 \\
\hline
3 & 128 & 256 & (5, 3) & (2, 1) & 1 \\
\hline
4 & 256 & 512 & (5, 4) & (2, 2) & 1 \\
\hline
5 & 512 & 1 & (6, 3) & (3, 2) & 0 \\
\hline
\end{tabular}
\end{table}

During the forward pass of the discriminator, the input image and the generated image are concatenated along the channel dimension. The concatenated image is then passed through each block layer. The sigmoid activation function is applied to the output of the patch layer to obtain probabilities between 0 and 1, producing a probability map that indicates whether the input image is real or generated.

\end{document}